\documentstyle[12pt,epsfig]{article}
\newlength{\dinwidth}
\newlength{\dinmargin}
\newcommand{\ba}{\begin{array}}
\newcommand{\ea}{\end{array}}
\newcommand{\be}{\begin{equation}}
\newcommand{\ee}{\end{equation}}
\newcommand{\bea}{\begin{eqnarray}}
\newcommand{\eea}{\end{eqnarray}}
\newcommand{\gsim}{\mathrel{\mathop{\kern 0pt \rlap
  {\raise.2ex\hbox{$>$}}} \lower.9ex\hbox{\kern-.190em $\sim$}}}

\def\nn{\nonumber}
\def\cO{{\cal O}}

\def\bu{{\bar{u}}}
\def\bw{{\bar{w}}}

\def\cO{{\cal{O}}}

\def\log{{\rm{log}}}

\def\cG{{\cal{G}}}

\begin{document}
\thispagestyle{empty}
\addtocounter{page}{-1}
\begin{flushright}
SNUST-000702\\
TIFR-TH/00-40\\
{\tt hep-th/0008042}
\end{flushright}
\vspace*{1.3cm}
\centerline{\large \bf Open Wilson Lines in Noncommutative Gauge Theory}
\vskip0.3cm
\centerline{\large \bf and}
\vskip0.3cm
\centerline{\large \bf Tomography of Holographic Dual Supergravity
~\footnote{Work supported in part by BK-21 Initiative in Physics (SNU - 
Project 2), KRF International Collaboration Grant, and KOSEF 
Basic Research Program 98-07-02-07-01-5 and 2000-1-11200-001-1.}}
\vspace*{1.2cm} 
\centerline{\bf Sumit R. Das${}^{a}$ {\rm and} Soo-Jong Rey${}^{b}$}
\vspace*{0.8cm}
\centerline{\it Tata Institute for Fundamental Research}
\vspace*{0.2cm}
\centerline{\it Homi Bhabha Road, Mumbai 400 005 INDIA ${}^a$}
\vspace*{0.6cm}
\centerline{\it 
School of Physics \& Center for Theoretical Physics}
\vspace*{0.2cm}
\centerline{\it Seoul National University, Seoul 151-742 KOREA ${}^b$}
\vspace*{1cm}
\centerline{\tt das@theory.tifr.res.in \hskip1cm sjrey@gravity.snu.ac.kr}
\vspace*{1.5cm}
\centerline{\bf abstract}
\vspace*{0.5cm}
We study the issue of gauge-invariant observables in 
$d=4, {\cal N} = 4$ noncommutative gauge theory and UV-IR relation
therein. We show that open Wilson lines form a complete set of gauge 
invariant operators, which are local in {\sl momentum} space and, depending
on their size, exhibit two distinct behaviors of the UV-IR relation. 
We next study these properties in a proposed dual description in terms of 
supergravity and find agreement.

\vspace*{1.1cm}

\baselineskip=18pt
\newpage

\section{Introduction}
Field theories defined on noncommutative spaces have attracted a lot
of attention recently \cite{ncstring}, particularly, 
because of the appearance of
noncommutative Yang-Mills (NCYM) theories as low energy limits of open
string theories in the background of NS-NS two-form potential $B^{\rm NS}$
\cite{sw}. A crucial consequence of
such noncommutativity is the growth of transverse size of objects with
increasing momentum \cite{sussk}. 
On the other hand, one expects that the large-N NCYM theory has a
description in terms of a supergravity dual when the 't Hooft coupling
$\lambda_{\rm eff} \equiv 4 \pi g_{\rm YM}^2 N$ is large. Such dual 
backgrounds with nonzero 2-form $B^{\rm NS}$ have been proposed 
\cite{hashitzhaki,maldarusso} and several aspects of the holographic map 
have been studied in \cite{ncholo}.  

In the absence of $B^{\rm NS}$ background, e.g. for the duality between
$N=4$ SYM theory in 3+1 dimensions and supergravity in $AdS_5 \times S^5$,
the scale in the YM theory is related to the radial coordinate in
$AdS_5$ \cite{wilsonloops, adsiruv}. 
For example, a source for some supergravity mode located in
the bulk of $AdS_5$ induces an expectation value of the dual operator
of the boundary Yang-Mills theory
\cite{banksbala}. If the background metric is chosen
to be
\begin{equation}
ds^2 = u^2 [-(dt)^2 + (dx^1)^2 + (dx^2)^2 + (dx^3)^2] + { (du)^2 \over u^2}
+ ( d \Omega_5)^2,
\label{eq:aone}
\end{equation}
then the latter expectation value has a support over a region of
size (in the $x_1 \cdots x_3$ directions) $ \sim ~{\cal O}(1/ \bu)$, where
$\bu$ refers to the location of the supergravity source. As the source
moves further away from (closer to) 
the boundary, the size of the corresponding 
disturbance in the boundary theory increases (decreases).

In NCYM theory, there are no local, gauge-invariant operators. However,
the theory has translation invariance. States in the theory are
still labelled by the energy and the momenta and hence there ought to be
operators in {\em momentum} space which create these states.
Consequently, if we place a source in the dual supergravity background
which has definite momentum along the $x^1 \cdots x^3$ directions and
a definite location in the remaining ``radial'' direction, this should
induce an expectation value of some operator in the gauge theory with
the same value of the momentum.

In this paper, we argue that the most natural basis of gauge-invariant 
operators of NCYM theory is provided by open Wilson line operators with
definite momentum, originally constructed in \cite{kawaiwilson} and identified 
with macroscopic fundamental
strings in \cite{reyunge}. We find that gauge invariance requires `size' of 
the Wilson line proportional to `momentum' along the noncommutative directions 
in a way consistent with the expected behavior in noncommutative theories. We
then consider the dual supergravity and study the hologram of a
source for a given supergravity mode by computing a one-point
correlation function.  Computation of correlation functions in such supergravity
backgrounds via evlauation of the supergravity action is generally ambiguous 
because of the necessity of momentum-dependent wave function
renormalizations. Our computation follows the unambiguous prescription
proposed in \cite{dasghosh}. From the momentum dependence of the
one-point correlators, we `tomograph' the profile of the hologram. When the
source is located deep inside the bulk, we show that the relationship
between the location of the source and the size of the hologram
approaches the standard relation found in the absence of the 
$B^{\rm NS}$-field : the
size of the hologram increases as the source moves further into the
bulk.  This is expected, since the region deep in the bulk corresponds
to the infrared regime of NCYM theory where the effects of
noncommutativity are invisible. However, when the source is located near the
boundary we find an opposite relationship : as the source moves closer
to the boundary, the size of its hologram {\em increases}.  The
relationship between the hologram size and location of the source is
found to be consistent with the relationship between the `size' and the
`momentum' of the open Wilson line operator in NCYM theory. Thus, the proposed
supergravity duals indeed encode the UV/IR relationship of NCYM
theory.

\section{Open Wilson Lines in Noncommutative Gauge Theory}
\subsection{Noncommutatitve Gauge Theory}
We will begin with a brief recapitulation of noncommutative gauge theory
and set notations. In this section, for simplicity, we will be 
considering gauge group U(1). Extension to U(N) group is straightforward
and only involves introduction of matrix-valued gauge fields. We start 
with definition of the generalized Moyal product:
\be
\phi_1(x) \star \phi_2 (y) \equiv
\exp \left( {i \over 2} \theta^{\mu \nu} \partial_\mu^x \partial_\nu^y
\right) \phi_1 (x) \phi_2 (y)
\label{moyal}
\ee
and Moyal commutator
\be
\left\{ \phi_1(x), \phi_2(y) \right\}_\star
\equiv \phi_1(x) \star \phi_2 (y) - \phi_2(y) \star \phi_1(x).
\ee
Throughout this paper, we will be studying `magnetic' noncommutativity:
$\theta^{23}:= \theta$ is the only nonvanishing component. In the context
of D3-brane in Type IIB string theory, the noncommutativity parameter is
determined by nonzero NS 2-form potential $B^{\rm NS}_{23}$:
\be
\theta^{\mu \nu} = \left( {1 \over B_{\rm NS}} \right)^{\mu \nu}
\qquad \quad {\rm or}, \quad {\rm equivalently}, \qquad \quad
\theta^{\mu \nu} B^{\rm NS}_{\nu \lambda} = \delta^\mu_\lambda. 
\ee 
Turning on 
$\theta$ breaks the underlying SO(3,1) Lorentz invariance to SO(1,1)  
Lorentz times SO(2) rotational invariance on the commutative and 
noncommutative subspaces, respectively \footnote{In what follows, we 
will denote the four-dimensional coordinates as $x, y, z \cdots$ and the
noncommutative subspace coordinates as ${\bf x}, {\bf y}, {\bf z} \cdots$.} 
and the commutative subspace coordinates as ${\bf x}_\perp, {\bf y}_\perp,
{\bf z}_\perp, \cdots$. For both commutative and noncommutative directions,
translational symmetry remains intact. As such, energy and momentum 
are conserved quantities and can be employed in labelling states and
operators.

Introduce U(1) gauge connection ${\bf A}_\mu(x)$ and define 
gauge-{\sl covariant} field strength ${\bf F}_{\mu \nu}(x)$ in terms of
the generalized Moyal product:
\be
{\bf F}_{\mu \nu} (x) \equiv
\left( \partial_\mu {\bf A}_\nu - \partial_\nu {\bf A}_\mu \right)(x)
+ \left\{ {\bf A}_\mu, {\bf A}_\nu \right\}_\star.
\ee
Gauge transformations of the noncommutative U(1) group is defined as
\bea
\delta_\epsilon {\bf A}_\mu(x) &=& \partial_\mu \epsilon(x) 
+ i \left\{ \epsilon, {\bf A}_\mu \right\}_\star (x)
\nonumber \\  
\delta_\epsilon {\bf F}_{\mu \nu} (x) &=& 
i \left\{ \epsilon, {\bf F}_{\mu \nu} \right\}_\star (x)
\label{gt1}
\eea
Neutral scalar fields ${\bf \Phi}^a(x)$ $(a=1, \cdots, 6)$ in `adjoint' 
representation transform similarly:
\be
\delta_\epsilon {\bf \Phi}^a (x) = i \left\{ \epsilon, {\bf \Phi}^a 
\right\}_\star (x).
\label{gt2}
\ee

Four-dimensional, noncommutative U(1) gauge theory arising from the 
low-energy worldvolume dynamics of a D3-brane in nonzero 
$B^{\rm NS}_{\mu \nu}$ background is then defined by  the following action:
\be
S = {1 \over 4 g^2} \int \! d^4 x \,\, 
\left(\, {\bf F}_{\mu \nu} \star {\bf F}^{\mu \nu}
+ D_\mu {\bf \Phi}^a \star D^\mu {\bf \Phi}^a + \left\{
{\bf \Phi}^a, {\bf \Phi}^b \right\}^2_\star \, \right).
\label{ncu1}
\ee

\subsection{Parisi's Composite Operators}
One distinguishing characteristic of noncommutative gauge theories is that
degrees of freedom in spacetime and in color space are all intertwined. 
This is rather obvious from the following simple observation. In rewriting 
a conventional gauge theory defined on noncommutative spacetime as a 
noncommutative gauge theory on commutative spacetime, one transmutes the 
{\sl color} degrees of freedom into the {\sl spacetime} degrees of freedom 
along the noncommutative directions. Thus, local observables of the form 
Tr$\, \hat{\cal O}$ in the former theory are now mapped into highly non-local 
ones of the form $\int d{\bf z} \, \hat{\cal O}(x)$ in the latter, 
where ${\bf z} \subset x$ refers to coordinates along the noncommutative 
directions. From this observation, it follows that, in general, it is 
impossible to define local operators in noncommutative gauge theories 
\footnote{By the same argument, it also follows that the
conventional notion of the operator product expansion or of the multipole
expansion does not make sense in noncommutative field theories.} .

If there are no gauge-invariant operators except the ones integrated
over the entire spacetime carrying zero energy and momentum, how can one even 
probe low-momentum, low-energy excitations? Actually, there is a class of 
gauge-invariant operators, which are sort of {\sl semi-localized} and hence 
may be used for probing the
noncommutative gauge theory excitations. These so-called Parisi operators
\cite{parisi}
carry definite energy and momentum, which are good quantum numbers in 
noncommutative spacetime, and are defined by Fourier modulation of 
a string of an elementary fields, $\phi_k (x)$, $(k = 1, 2, \cdots)$:
\be
{\cal O}_n (x_1, x_2, \cdots, x_n; {\bf k})
\equiv 
\int \! d^2{\bf z} \,\, \phi_1 ({\bf z} + x_1) \star \phi_2({\bf z} + x_2) \star 
\cdots \star \phi_n ({\bf z} + x_n) \star e^{ i {\bf k} \cdot {\bf z} } .
\label{nonlocalop}
\ee
Being integrated over noncommutative coordinates of all elementary fields, 
the Parisi operators are non-local. As viewed from momentum space, however, 
they are {\sl local} operators. Thus, we take a viewpoint that physically 
relevant operators in noncommutative field theory are the ones which are 
local in configuration space for commutative directions but in 
momentum space for noncommutative directions. Note also that, at this
stage, the multi-locations $(x_1, \cdots, x_n)$ are not directly related
to the momentum vector ${\bf k}$ along the noncommutative directions. 
Thus, in a noncommutative field theory, a class of physically relevant
$m$-point correlation functions for probing the theory is provided by:
\be
G_m ({\bf k}_1, {\bf k}_2, \cdots, {\bf k}_m) 
= \Big\langle {\cal O}_1 ({\bf k}_1) {\cal O}_2 ({\bf k}_2) \cdots
{\cal O}_m ({\bf k}_m) \Big\rangle,
\ee
where, for clarity of the definition, dependence on coordinates
$x_1, x_2, \cdots$ in each operators and in the resulting correlation 
functions are omitted.

Let us illustrate the utility and meaning of the Parisi operators in the 
simplest context: noncommutative scalar field theory. 
Take the one-point correlator:
\be
G_1 (x, {\bf k}) = \Big<
\int \! d^2{\bf z} \,\, \phi({\bf z}) \star \phi({\bf z} + x) \star 
e^{ i {\bf k} \cdot  {\bf z}} \Big>.
\ee
In order to visualize the spacetime picture, consider first the limit 
$x \rightarrow 0$. One finds that the Parisi operator reduces to
 a sort of $[\phi^2]$ composite operator, partially Fourier-transformed 
in the noncommutative directions, except that now
all the products (including Fourier transform) are defined in terms of
the Moyal product. Indeed, the above one-point correlator 
may be understood as follows. 
Introduce, in the noncommutative scalar field
theory Lagrangian, a bilocal mass `spurion' term:
\be
{\cal S}_{m^2} = {1 \over 2} \int d^4 x \!\! \int d^4 y \,\,\, m^2 (x, y) 
\star  \left[ \phi(x) \star \phi(y) + \phi(y) \star \phi(x) \right],
\ee
where all the products are the generalized Moyal product, Eq.(\ref{moyal}). 
Consider the situation that $x = x, y
= x + {\bf z}$, viz. non-local split only within the noncommutative subspace. 
Take the suprion mass-squared a slowly varying function in 
$x$ and Fourier expand:
\be
m^2 (x, {\bf z}) \, = \, \int \! {d^2{\bf k} \over (2 \pi)^2} \,\,
e^{i {\bf k} \cdot {\bf  z}} \, \widetilde{m^2} (x, {\bf k})
.
\ee
One then immediately find that the one-point correlators can be
derived from the partition function of the noncommutative scalar 
field theory:
\be
G_1 (x, {\bf k}) \,\,\, = \,\,\, {\delta \ln Z_{\rm NC}[\widetilde{m^2}] \over 
\delta \widetilde{m^2} (x, {\bf k}) } .
\ee
Extending the result, it should be fairly obvious that generic 
multi-point correlators involving the non-local operators of the type 
Eq.(\ref{nonlocalop}) are derivable from variation of the 
partition function with respect to a set of suitable spurion couplings. 

\subsection{Gauge Invariance = Spacetime Translation Invariance}
As mentioned above, in noncommutative field theories, degrees of freedom
in spacetime and in internal space are all intertwined. In particular, this 
has implied, in noncommutative gauge theory, there is {\sl no} gauge 
invariant, local observables. In view of its profound implication, in this
subsection, we would like to understand better the 
meaning of the noncommutative
gauge invariance. Indeed, we will be showing explicitly that noncommutative 
gauge invariance is identical to spacetime translational invariance and hence
vast reduction of the degrees of freedom -- intimately related to the 
Eguchi and Kawai reduction \cite{ek}. In fact, noncommutative gauge
theories can be derived from twisted version of the Eguchi-Kawai 
models \cite{kawaigroup}.

To grasp the physical meaning of the noncommutative gauge invariance, begin
with Fourier decomposition of the gauge parameter along the noncommutative 
directions:
\be
\epsilon({\bf x}) = \int \! {d^2 {\bf k} \over (2 \pi)^2} \,\, 
e^{ i {\bf k} \cdot {\bf x} } \,\, \widetilde{\epsilon}({\bf k}).
\ee
Dependence on the commutative coordinates are omitted for
notational simplicity.
One can then re-express the noncommutative gauge transformations 
Eqs.(\ref{gt1}, \ref{gt2}) as:
\bea
\delta_\epsilon {\bf A}_\mu ({\bf x}) &=& i \int {d^2 {\bf k} \over (2 \pi)^2}
e^{ i {\bf k} \cdot {\bf x}} \, \widetilde{\epsilon} ({\bf k}) \,
\Big[
\left( {\bf A}_\mu ( {\bf x} + \theta \cdot {\bf k}) - {\bf A}_\mu({\bf x}
- \theta \cdot {\bf k})
\right) + i \bf k_\mu \Big],
\nonumber \\
\delta_\epsilon {\bf \Phi}^a ({\bf x})
&=& i \int {d^2 {\bf k} \over (2 \pi)^2} \, e^{i {\bf k} \cdot {\bf x}} \, 
\widetilde{\epsilon} ({\bf k}) \, 
\Big[ {\bf \Phi}^a ({\bf x} + \theta \cdot {\bf k}) - {\bf \Phi}^a
({\bf x} - \theta \cdot {\bf k}) \Big].
\label{mmgt}
\eea
We recognize that the noncommutative gauge transformation is identified with 
the spacetime translation (plus an additional constant shift in case
of the gauge fields). Take $\widetilde{\epsilon}({\bf k})$ to be a Gaussian
distribution with dispersion $\Delta {\bf k}$. Then, Eq.(\ref{mmgt})
implies that dispersion of the spacetime translation $\Delta {\bf x}$ is 
given by
\be
\Delta {\bf x}^\mu \,\,\,\, \sim \,\,\,\, \theta^{\mu \nu} \Delta {\bf k}_\nu,
\label{uvirgauge}
\ee
viz. size of the spacetime translation is proportional to the Fourier wave
vector of the gauge transformation. 

To construct gauge invariant observables, one needs to average over the gauge 
orbits. According to the above interpretation, 
such observables are the ones integrated over the 
entire spacetime, as the gauge orbits are identified with orbits of the 
spacetime translation. While this is quite true and is intimately related to
the Eguchi-Kawai reduction, we show below that there exists a class of 
gauge-invariant operators in noncommutative gauge theory, open Wilson
line operators. In fact, they turn out precisely the gauge theory counterpart 
of the Parisi's composite operators discussed in the last subsection.  

\subsection{Open Wilson Line Operators}
\begin{figure}[ht]
   \vspace{0.5cm}
\centerline{
   {\epsfxsize=5.5cm
   \epsfysize=5cm
   \epsffile{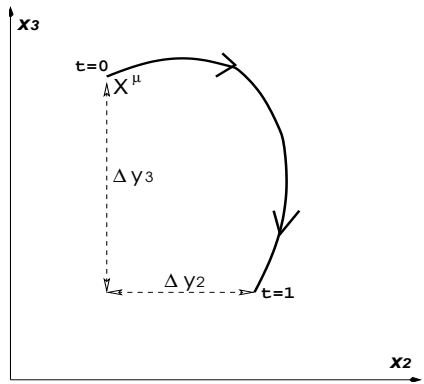}}
}
\caption{\sl Projection of the open Wilson line on the noncommutative
plane. The base-point at $t = 0$ is denoted as $x^\mu$, splitting 
between the two endpoints ($t=0, 1$) as $\Delta y^{2,3}$. }
\label{wilson}
\end{figure}

In noncommutative gauge theory, there is a distinguished class of such
{\sl semi-local} operators, which are labelled by momentum along the 
noncommutative directions -- {\sl open} Wilson lines: 
\be
W_{\bf k} [C] = \int \! d^2{\bf x} \, {\cal P} \, \exp_\star \left[ i \int_C
\left( \dot y(t) \cdot {\bf A} (x + y(t)) +
\sqrt{\dot{y}^2(t)} \hat\Omega (t) \cdot {\bf \Phi} (x + y(t)) \right) \right] 
\star e^{i {\bf k} \cdot {\bf x}}.
\label{openwilson}
\ee
This is a generalization of the operator constructed 
in \cite{kawaiwilson}.
Here, $t = [0, 1]$ denotes an affine parameter along a contour $C$ 
specifying the open  Wilson line and ${\cal P}$ denotes the path ordering
along $C$. The spacetime position of the base point $(t = 0$)
is denoted as $x^\mu$, which may also be viewed as encoding center-of-mass
position of the Wilson line. Similarly, along the Wilson line, the spacetime 
image of the point $t$ as measured relative to $x^\mu$ is denoted as
$y^\mu(t)$. Distance between the two endpoints of the open Wilson line 
is given by:
\be
\Delta y^\mu \equiv y^\mu (1) - y^\mu (0).
\ee
$\hat\Omega^a(t)$ ($a=1,2, \cdots, 6)$ refers to the angular coordinates on 
$S_5$ in Eq.(\ref{eq:aone}) having unit modulus, $\hat \Omega (t) \cdot 
\hat\Omega (t) 
= 1$.  For simplicity, in this paper, we will be studying open Wilson lines
consisting only of the gauge fields. 
A cartoon view of the open Wilson line is depicted in Fig.\ref{wilson}

Let us now examine explicitly under what conditions the open Wilson lines 
would be gauge invariant. Interestingly enough, we will be discovering that
the momentum ${\bf k}_\mu$ is not arbitrary but is directly related to
the splitting $\Delta y^\mu$. Denote finite gauge transformation parameter 
as U$(x) = \exp_\star (i \epsilon (x))$. 
First, under a gauge transformation that depends only on the commutative 
coordinates, the open Wilson line is trivially invariant, as the two
ends of the Wilson line are splitted only along the noncommutative 
directions. Thus, let us focus on the case where the gauge transformation
parameter depends on the noncommutative coordinates. We then find that
\be
W_{\bf k}[C] \quad \rightarrow \quad W^{\rm U}_{\rm k}[C] = 
\int \! d^2{\bf x} \, {\cal P} \, {\rm U} ({\bf x} + \Delta y) \star
e^{ i \int \dot y \cdot {\bf A}(x + y)} \star {\rm U}^{-1} ({\bf x}) 
\star e^{ i {\bf k} \cdot {\bf x}}.
\label{gaugetransf}
\ee
From the definition of the Moyal product, it follows that
\be
e^{+ i {\bf k} \cdot {\bf x}} \star {\rm F}({\bf x} + \theta \cdot {\bf k})
\star e^{- i {\bf k} \cdot {\bf x}} = {\rm F}({\bf x})
\qquad
{\rm and}
\qquad
e^{- i {\bf k} \cdot {\bf x}} \star {\rm F}({\bf x})
\star e^{+ i {\bf k} \cdot {\bf x}} = {\rm F}({\bf x} + \theta \cdot {\bf k})
\ee
for any function F$({\bf x})$ and hence
\be
e^{i {\bf k} \cdot {\bf x}} \star {\rm U}^{-1}
({\bf x} + \theta \cdot {\bf k})
= {\rm U}^{-1} ({\bf x}) \star e^{i {\bf k} \cdot {\bf x}}.
\label{relation}
\ee
This is nothing but the Moyal product manifestation of the result 
shown in the last subsection 
that noncommutative gauge transformation is equivalent to translation 
along noncommutative directions. 

Applying Eq.(\ref{relation}) to the last expression in Eq.(\ref{gaugetransf})
and using the cyclic property of the Moyal product, we finally find that
\bea
W^{\rm U}_{\bf k}[C] &=& \int \!\! d^2 {\bf x} \,\, {\cal P} \, 
{\rm U}({\bf x} + \Delta y) \star e^{ i \int \dot y \cdot {\bf A}(x + y) } 
\star e^{i {\bf k}\cdot {\bf x}} \star {\rm U}^{-1} ({\bf x} + \theta 
\cdot {\bf k})
\nonumber \\
&=&
\int d^2 {\bf x} \,\, {\cal P} \, {\rm U}^{-1} ({\bf x} + \theta \cdot {\bf k})
\star {\rm U}({\bf x} + \Delta y) \star  e^{ i \int \dot y \cdot {\bf A}(x + y) } \star e^{i {\bf k} \cdot {\bf x}}
\nonumber \\
 &=& W_{\bf k} [C]
\eea
{\sl provided} the splitting $\Delta y$ is nonzero only along the 
noncommutativity directions and is related to the momentum ${\bf k}$
of the open Wilson line as:  
\be
{\bf k}_\mu  \,\,\, = \,\,\, \left( {1 \over \theta} \right)_{\mu \nu} 
\Delta y^\nu \, .
\label{uvir}
\ee
Thus, we have proven that the open Wilson lines Eq.(\ref{openwilson}) are 
indeed gauge invariant but only under the condition that the `momentum' 
${\bf k}$ is related to the `size' $\Delta y$ precisely as in Eq.(\ref{uvir}). 
Along with Eq.(\ref{uvirgauge}), the relation Eq.(\ref{uvir}) stems from 
the underlying
noncommutative gauge invariance. As such, we take Eq.(\ref{uvir}) as the
fundamental defining relation characterizing all gauge invariant open Wilson 
lines. 

One may wonder how possibly the above Wilson lines can be encoded into
the noncommutative gauge theory. The Parisi's prescription alluded above
tells us what ought to be done. In the gauge theory partition function, 
consider a `naive' open Wilson line along a contour $C$ and couple it
to a non-local `spurion' field:
\be
{\cal S}_J = \int \! d^4 y \int \!d^4 x \,\, {\cal P} \, 
\exp_\star \left( i \int_C \dot y(t) \cdot {\bf A} (x + y (t) \right) 
\star J [ x , C(y)].
\label{wlspurion}
\ee
Note that the naive Wilson line is {\sl not} gauge invariant. Therefore,
under the noncommutative U(1) gauge transformation,
the spurion coupling $J[x, C(y)]$ ought to transform appropriately so 
that $S_J$ is rendered gauge invariant. Fortuitously, due to the 
equivalence of the gauge transformation and the spacetime translation, 
it is possible to 
extract the gauge variant piece out of the `spurion' field and adjoin it
to the naive Wilson line. To do so, consider the `spurion' field slowly 
varying along the commutative directions. One can partially Fourier-expand 
it along the noncommutative direction as:
\be
J [x, C(y)] = \int {d^2 {\bf k} \over (2 \pi)^2}
\,\, e^{ i {\bf k} \cdot {\bf x}} \, \widetilde{J}_{\bf k} 
[{\bf x}_\perp, C(y)].
\ee
Inserting this to Eq.(\ref{wlspurion}), one finds
\be
{\cal S}_J = \int d^4 y \int d^2{\bf x}_\perp \int {d^2 {\bf k}
\over (2 \pi)^2} \,\, W_{\bf k} [{\bf x}_\perp, C(y)]
\, \widetilde{J}_{\bf k} [{\bf x}_\perp, C(y)].
\label{wlspurion2}
\ee
Here, $W_{\bf k}[C]$ is indeed the gauge invariant open Wilson line.
Thus, even though $J[x, C]$ is not gauge invariant, Fourier-transformed
`spurion' coupling $\widetilde{J}_{\bf k}[{\bf x}_\perp, C]$ 
has become gauge invariant. 
Moreover, 
in the integrand, the open Wilson line operator and the source
$\widetilde{J}_{\bf k} [{\bf x}_\perp, C]$ are multiplied 
as an ordinary product, as they
are functions of the commutative coordinates only. 
As such, one-point correlator of the gauge invariant open Wilson line can
be obtained from functional response of the gauge theory with respect to
the {\sl Fourier-transformed} `spurion' couplng:
\be
\Big< W_{\bf k} [{\bf x}_\perp, C(y)] \Big>_{\rm NCYM} \quad
\quad = 
\qquad {\delta \ln Z_{\rm NCYM}[\widetilde{J}] 
\over \delta \widetilde{J}_{\bf k} [{\bf x}_\perp,
C(y)]}.
\ee

Incidentally, the spurion coupling to the Wilson lines also 
indicates an interesting reshuffling between the noncommutative 
coordinates. Namely, even though the coupling
Eq.(\ref{wlspurion}) involves sum over size and base point location of
the open Wilson lines, the more natural, gauge-invariant coupling 
Eq.(\ref{wlspurion2}) indicates that the sum ought to 
be interpreted in terms of the commutative coordinates ${\bf x}_\perp$
and momentum ${\bf k}$ along noncommutative directions. 

Lastly, much as the closed Wilson loops form a complete set of gauge invariant
observables in conventional gauge theory, we can take the open Wilson lines
Eq.(\ref{openwilson}) as a complete set of gauge invariant
observables in the noncommutative gauge theory.

\subsection{UV-IR Relation and Spacetime Uncertainly Principle}
\begin{figure}[ht]
   \vspace{0.5cm}
\centerline{
   {\epsfxsize=10.5cm
   \epsfysize=3.5cm
   \epsffile{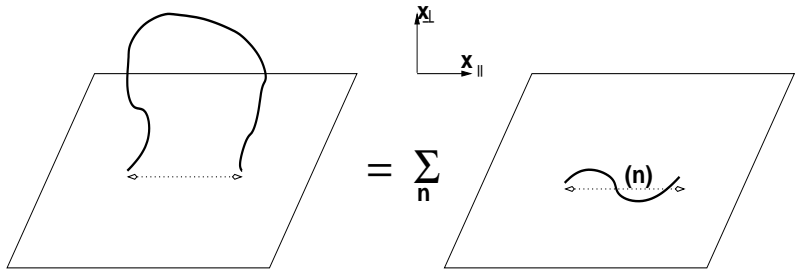}}
}
\caption{\sl Multipole expansion of open Wilson lines.}
\label{expansion}
\end{figure}

The relation Eq.(\ref{uvir}) is the most important, salient feature 
of the noncommutative Wilson lines in that it relates the `momentum'
 and the `size' 
of the Wilson line. It should be emphasized that, by assigning the Fourier 
mode ${\bf k}$, two spacetime aspects of the Wilson line are specified --- 
the momentum of the base point $x$ and the size of the open Wilson line 
$\Delta y$. Typical open Wilson lines have a shape whose
endpoints are splitted within the noncommutative subspace but not in 
commutative subspace. 
While the standard notion of the operator product expansion does not make
sense along the noncommutative directions, it still
holds along the commutative directions. Recalling that the Wilson line does 
not have any endpoint splitting in the commutative subspace, it would
make a sense to expand a generic open Wilson line in the ambient spacetime
into multipoles of a subset of open Wilson lines which lie entirely within
the noncommutative subspace. This is illustrated in Fig.(\ref{expansion}).

To grasp the physical meaning, we shall be exploring the Wilson line 
operators at two different regimes of the spatial momentum, ${\bf k}$. 

First, let us consider probing physics at `infrared' as compared to the 
noncommutativity scale, $\vert {\bf k} \vert \ll 1/\sqrt{\theta}$. The Moyal 
phase factors in Eq.(\ref{moyal}) are completely suppressed that all the Moyal 
products are reduced to ordinary products. This implies that, in the 
infrared regime, the noncommutative gauge theory converges to the conventional 
commutative gauge theory \footnote{ This reduction
remains the same even at quantum level, as the underlying ${\cal N}=4$
supersymmetry ensures absence of ultraviolet divergences in the noncommutative
gauge theory and hence no possibility of UV/IR mixing. }.
In this regime, $\vert {\bf k} \vert \ll 1/\sqrt{\theta}$ and Eq.(\ref{uvir})
indicates that $\vert \Delta y \vert \ll \sqrt{\theta}$, viz. `size' of 
the open Wilson line is smaller than the noncommutativity scale,
a minimum distance scale one can probe. Altogether with vanishing
Moyal phase factor, the Wilson line operator is also reduced effectively to a 
Fourier-transform of the standard, commutative, {\sl closed}
Wilson loop operator. Therefore, in the infrared regime, the conventional 
${\cal N}=4$ supersymmetric gauge theory would describe dynamics well and 
the well-known UV-IR {\sl duality} \cite{wilsonloops}
would follow immediately at large-N and strong `t Hooft coupling limit. 

Incidentally, the closed Wilson loop operators are known
to form a complete set of gauge-invariant physical observables in the
conventional, commutative gauge theories. Uniqueness of the Moyal 
deformation (up to gauge equivalence) as a noncommutative but associative 
deformation then implies that the open Wilson line operators in noncommutative 
gauge theory also form a complete set of gauge invariant observables. Hence,
together with the fact that they can carry nonzero momentum along the
noncommutative directions, we conclude that the open Wilson line operators 
are the most physically suitable probes for dynamical aspects of the 
noncommutative gauge theories.

Next, consider probing physics at `ultraviolet' as compared to the 
noncommutativity scale, $\vert {\bf k} \vert \gg 1/\sqrt{\theta}$. From 
Eq.(\ref{uvir}), one finds that the `size' of the open Wilson line would 
be larger than the noncommutativity scale,
$\vert \Delta y \vert \gg \sqrt{\theta}$. In fact, these excitations look
like macroscopically large fundamental open strings stretched along 
the noncommutativity subspace. 
As such, in the `ultraviolet' regime, we would find UV-IR {\sl proportionality}
instead of {\sl duality}. Note that we have reached such a conclusion purely
based on noncommutative gauge theory --- open Wilson lines are the only
known operators with gauge invariance and nonzero momentum along the 
noncommutative directions.  

A remarkable consequence of Eq.(\ref{uvir}) is that the open Wilson lines
satisfy a version of (Euclidean) spacetime uncertainty relation \cite{yoneya}. 
If the 
macroscopic open Wilson lines are treated quantum mechanically, ${\bf k}_{1,2} \sim 
\Delta {\bf k}_{1,2} \ge \hbar/\Delta x^{1,2}$, where $x^\mu$ denotes the coordinates
of the open Wilson line base point at $t = 0$. Thus, utilizing the UV-IR
relation Eq.(\ref{uvir}), one immediately finds
\be
\Delta x^1 \Delta y^2 \gsim {1 \over 2} \hbar \theta
\qquad \quad
{\rm and} \qquad \quad
\Delta y^1 \Delta x^2 \gsim {1 \over 2} \hbar \theta. 
\ee
Here, recall that $\Delta y$ refers to the `classical' distance between
the two endpoints of the open Wilson line. As such, the above inequalities
define a version of (Euclidean) spacetime uncertainty relation   
satisfied between the classical size and the quantum mechanical uncertainty
of the base point position of the open Wilson lines.
In other words, base-point coordinates and the open Wilson line
splitting distances form a set of conjugate variables each other. 
 
\section{Tomography of Five-Dimensional Supergravity Dual}
We now turn to supergravity description of the noncommutative gauge theories,
as initiated in \cite{hashitzhaki} and \cite{maldarusso}.
As in the previous sections, we shall be considering scaling limit of 
D3 branes (oriented along $x^1 \cdots x^3$ directions) in the presence of 
$B_{23}^{\rm NS} \neq 0$. The resulting background is described by a classical 
solution of the Type IIB supergravity, whose string frame metric is given by
\be
ds^2_{\rm IIB} = \alpha ' R^2 \, 
\left[u^2 (- (dx^0)^2 + (dx^1)^2) + {u^2 \over 1+ a^4 u^4}
((dx^2)^2 + (dx^3)^2) + {du^2 \over u^2} +d\Omega_5^2 \right],
\label{eq:bone}
\ee
while the dilaton $\phi$, NS-NS 2-form potential $B_2^{\rm NS}$ and 
the R-R 2-form potential $C_2^{\rm RR}$ and the R-R self-dual, 5-form 
field strength $F_5^{\rm RR}$ are given by
\bea
e^{2\phi} & = & g^2 {1 \over 1+ a^4 u^4} \,\, \quad \,\qquad B^{\rm NS}_{23}
= {\alpha ' R^2 \over a^2}{a^4u^4 \over 1+ a^4 u^4} \nn \\
 C^{\rm RR}_{01} & = & {1 \over g} {\alpha '  R^2 \over a^2} a^4 u^4
\qquad \quad F^{\rm RR}_{0123u}
= {\alpha '^2 \over g} {1 \over 1+ a^4u^4}\partial_u (u^4 R^4) .
\label{eq:bthree}
\eea
Here, $R^4 = 4\pi gN$ and $g$ refers to the {\it open} string coupling.
In the infrared, $ua \ll 1$, $B^{\rm NS}_2, C^{\rm RR}_2$ tend
to vanish and the spacetime asymptotes to $AdS_5 \times S^5$, the supergravity
dual to the large-N limit of the standard $d=4, {\cal N}=4$ supersymmetric
gauge theory.

In this background, the graviton fluctuation $h_{01}(x, u)$ with 
zero momenta along $x^0, x^1$ and zero angular momenta along the
$S^5$ satisfy a simple decoupled equation.
Denoting $\phi (x, u) = g^{00}h_{01} = g^{11} h_{01}$ and expanding the 
Type IIB supergravity action, one easily finds that the $\phi$-field equation
in string frame is given by
\be
\partial_\mu(\sqrt g e^{-2\phi} g^{\mu\nu}\partial_\nu
~\phi) = 0. 
\label{eq:bfour}
\ee
In terms of Fourier modes along the noncommutative directions
\be
\phi (u, {\bf x}) = \int {d^2 {\bf k} \over (2\pi)^2} 
\, e^{-i{\bf k} \cdot {\bf x}} \,\widetilde\phi ( {\bf k},u) ,
\label{eq:bfive}
\ee
the field equation becomes
\be
\partial_u \Big(u^5 \partial_u \widetilde\phi ({\bf k},u) \Big)
- {\bf k}^2 u \Big(1 + a^4 u^4 \Big)\, \widetilde\phi ({\bf k},u) = 0 .
\label{eq:bsix}
\ee
Here, ${\bf k}^2 = k_2^2 + k_3^2$.
Eq.(\ref{eq:bsix}) represents a perturbation of the background 
with zero-energy. As such, it does
not make sense in Lorentzian signature. Hence, in what follows, we will 
work always in the Euclidean signature.

The background Eqs.(\ref{eq:bone}, \ref{eq:bthree}) has been proposed as the
holographic dual of the noncommutative gauge theory residing at the boundary, 
$u = \infty$. Noting that the value of $B^{\rm NS}_{23}$ at the
boundary is $B_{23}^{\rm NS} = {\alpha ' R^2 / a^2}$, 
 noncommutativity scale in the gauge theory is identified as 
$\sqrt{\theta} \equiv \sqrt{\alpha ' / B^{\rm NS}_{23}} =  {a / R }$. 
In the supergravity background, however, the spacetime metric 
Eq.(\ref{eq:bone})
indicates that the scale of departure from $AdS_5 \times S^5$ is really
$a$ rather than $a/R$. This difference may be attributed to strong
coupling dynamics of the noncommutative gauge theory, much as in the 
standard AdS/CFT correspondence, where similar non-analytic enhancement 
has been discovered \cite{wilsonloops}. 

Our goal in this section would be to understand the UV/IR relation 
exhibited by the open Wilson line operators from
the dual supergravity side. In parallel to the gauge theory analysis
in the previous section, we will be focusing on supergravity counterpart
of the one-point correlators of gauge-invariant operators. For doing so, we
need to begin with a general prescription for the one-point correlator.
\subsection{One-Point Correlator from Dual Supergravity}
From the supergravity side, to study one-point correlator, we add a suitable
nondynamical source to the graviton fluctuation, $\phi$, in 
Eq.(\ref{eq:bfour}). As in the previous section, we assume the source is
prescribed by a definite momentum ${\bf k}$ in the noncommutative 
directions, $(x^2,x^3)$, and by a definite location ${\bar u}$ in the radial 
direction. In the dual gauge theory, turning on the source simply refers to 
a situation that we have excited the system from its ground state. 
Thus, in the presence of the source, the operator $\cO ({\bf k})$ dual to 
the $\phi$-field would be acquiring a nonvanishing vacuum expectation value. 
We are interested in 
`tomography' of the one-point correlator --- profile of  
$\langle \cO ({\bf x}) \rangle_\bu$ as a function of ${\bf x}$ and $\bu$.

In the standard AdS/CFT correspondence, where the NS-NS 2-form potential 
$B^{\rm NS}$ is turned off, prescription for the one-point correlator is 
well understood --- perturb the supergravity field around the normalizable 
solution $\phi_{\rm norm}$ of Eq.(\ref{eq:bsix}) in the presence of a source:
\be
\phi ({\bf k},u) \,\, = \,\, \phi_{\rm norm}({\bf k}, u) + 
\delta \phi_{\rm non-norm} ({\bf k}, \bu). 
\nonumber
\ee
Here, $\delta \phi_{\rm non-norm}({\bf k}, \bu)$ 
denotes a suitable non-normalizable 
mode. Evaluating the supergravity action for $\phi ({\bf k}, u)$ and
taking functional variation of the action with respect to $\delta 
\phi_{\rm non-norm} ({\bf k}, \bu)$ would then yield the one-point correlator,
$\langle \cO ({\bf k}) \rangle_\bu$.

In the presence of the $B^{\rm NS}$ background, a similar procedure was 
carried out to calculate one-point correlator in the presence of a D-instanton
in \cite{drt}. There, it was found that there is an ambiguity in 
extracting the one-point correlators due to necessity of momentum-dependent 
wave-function renormalizations. A similar ambiguity was found for two-point 
correlators in \cite{maldarusso}. 

For two-point correlators, a possible resolution \footnote{
A closely related prescription was proposed in 
\cite{danielsson} in the context of holography in the full D3-brane 
geometry.} of the ambiguity 
has been suggested in \cite{dasghosh}. The idea of \cite{dasghosh}
was to postulate an operator-field correspondence between the gauge theory 
and the dual supergravity along the lines of \cite{banksbala}. After 
Wick-rotation to Euclidean signature, it then followed that the
two-point correlator of a gauge theory operator $\cO$ ought to be given by
\be
\Big< \cO ({\bf k})\cO(-{\bf k}) \Big>_{\rm E}
= {\rm Lim}_{u,u' \rightarrow \infty} \left({{\bf k}^4
\over \Psi_{{\bf k}}(u) \Psi_{-{\bf k}}(u')} \right)
\Big[ \cG_{\rm E} (u,u'; {\bf k}) - \cG_0(u,u'; {\bf k}) \Big].
\label{eq:bseven}
\ee
Here, $\cG_{\rm E} (u,u'; {\bf k})$ denotes the Euclidean {\em bulk} Green 
function in the supergravity background Eqs.(\ref{eq:bone}, \ref{eq:bthree}), 
$\cG_0 (u,u'; {\bf k})$ is the bulk Green function in flat space,
and $\Psi_{\bf k}(u)$ is the (Wick-rotated) normalized wave function pertinent
to the perturbation. Clearly, the normalizations in all the
quantities in Eq.(\ref{eq:bseven}) are well defined such that we have
an unambiguous supergravity prediction for the gauge theory correlator.
It was shown in \cite{dasghosh} that the correlator Eq. (\ref{eq:bseven})
has the correct low-$(ka)$ behavior.

Here, we utilize the same prescription for the one-point correlator in the
presence of a source located at $\bu$ in the radial direction. Then,
the classical solution of $\phi$ at a radial position $u$ is simply given by 
the bulk Green function $\cG_{\rm E} (u,\bu; {\bf k})$. 
Following the same steps as in \cite{dasghosh}, we finally get
\be
\Big< \cO({\bf k}) \Big>_{\bu} \,\, = \,\, {\rm Lim}_{u \rightarrow \infty} 
~\left( { {\bf k}^2 \over \Psi_{\bf k}(u)} \right)~ \Big[
\cG_{\rm E} (u,\bu; {\bf k}) - \cG_0 (u,\bu; {\bf k}) \Big].
\label{eq:beight}
\ee
Provided the limit specified does exist and yields a $u$-independent result,
Eq.(\ref{eq:beight}) may be taken as {\sl the} defintion of 
the one-point correlators from the dual supergravity side.  
\subsection{Consistency Check -- AdS/CFT Correspondence}
One can easily confirm that the prescription Eq.(\ref{eq:beight}) yields the 
correct result for $AdS_5 \times S^5$ background. In this case, the
Euclidean bulk Green function for a massless scalar field is given by
\be
\cG_{\rm E} (u,\bu; k) \Big\vert_{\rm AdS}
\,\, = \,\, \left({1 \over u\bu} \right)^{2}~K_2 \left({ k \over \bu} \right)~
I_2 \left({k \over u} \right)
\qquad \quad {\rm for} \qquad \quad u > \bu.
\label{eq:bnine}
\ee
Here, $k$ refers to the magnitude of the Euclidean 4-momentum along the 
D3-brane worldvolume directions. After Wick rotation to the Euclidean 
signature, the orthonormal wave function is given by $\Psi_{\bf k}(u) 
u^{-2} I_2 (k/u)$. Moreover, for a fixed $\bu$, as $u \rightarrow \infty$,
the bulk Green function in {\sl flat} space
\be 
\cG_0 (u,\bu;k) \,\, = \,\, 
{1\over 2ka^2} {1\over (u\bu)^{5/2}} \exp \left( -ka^2 \left( u - \bu 
\right) \right)
\label{eq:bninea}
\ee
goes to zero exponentially, while the wave function $\Psi_{\bf k}(u)$ 
goes to zero only in powers of $1/u$.
Thus, as $u \rightarrow \infty$ limit is taken, the subtraction of
$\cG_0 (u, \bu; k)$ 
in Eq.(\ref{eq:beight}) is irrelevant for $AdS_5 \times S^5$
background. One obtains
\be
\Big< \cO(k) \Big>_\bu \, \Big\vert_{\rm AdS} \,\, = \,\, 
\left( {k \over \bu} \right)^2 ~K_2 \left({k \over \bu} \right).
\label{eq:bten}
\ee
The `tomograph' of the source distributed on the holographic boundary
at $u  = \infty$ is then provided by Fourier-transform of Eq.(\ref{eq:bten}):
\bea
\Big<\cO(\Delta x) \Big>_{\bu} & = & \int {d^4 k \over (2\pi)^4}
e^{-ik \cdot \Delta x}~ \left({ k \over \bu } \right)^2~
K_2 \left({ k \over \bu} \right) \nn \\
& = & {\bu^{-4} \over \left[(\Delta x)^2 + \bu^{-2} \right]^4}.
\label{eq:beleven}
\eea
Thus, as the bulk source approaches the boundary, $\bu \rightarrow  \infty$,
the tomograph of $\langle\cO(\Delta x) \rangle_{\bu}$ asymptotes to
Dirac delta function, $\delta^{(4)} (\Delta x)$. As such, characteristic
size of the hologram may be extracted from equal-altitude contours of
$\langle \cO(\Delta x) \rangle_\bu$ being a constant multiple of $\bu^4$. 
It yields
\be
\Big( |\Delta x| \Big)_0 \quad \sim \quad  {1\over \bu}
\qquad \quad {\rm for} \quad {\rm all} \quad \bu,
\label{eq:btwelve}
\ee
increasing (decreasing) {\sl monotonically} as the source moves into 
(out of) the bulk.
This is a manifestation of the UV/IR-{\sl duality} \cite{wilsonloops}
of the AdS/CFT correspondence reflected in the one-point correlator.

Incidentally, it is also instructive to understand the relationship 
Eq.(\ref{eq:btwelve}) qualitatively for small $\bu$ regime. 
In this regime, $K_2 (k / \bu) \, \sim  \, 
\sqrt{\bu} \exp( - k/ \bu )$ and the integral over $k$ in
Eq.(\ref{eq:beleven}) leads to a vanishing contribution for 
$\Delta x \gg 1/\bu$.

\subsection{Noncommutative One-Point Correlator}
We finally study the one-point correlator Eq.(\ref{eq:beight}) in the
presence of $B^{\rm NS}_{23} \ne 0$. 
In \cite{drt, dasghosh}, the bulk Green function for a nonzero $B^{\rm NS}$
background has been computed. In Euclidean space, the result is:
\be
\cG_E (u,\bu; {\bf k}) = {\pi \over 4i} {C(ka) \over A(ka)}
{1 \over u^2 \bu^2}
H^{(1)} \left(\nu,\bw+{i\pi\over 2} \right)
H^{(2)} \left(\nu, -w-{i\pi\over 2} \right)
\qquad {\rm for} \qquad u > \bu,
\label{eq:bthirteeen}
\ee
where, as in the previous sections, 
$\bf k$ is the momentum vector along the $(x^2, x^3)$ noncommutative 
directions, $k$ is its magnitude \footnote{The bulk Green function 
Eq.(\ref{eq:bthirteeen}) is valid 
also for nonzero Euclidean momentum ${\bf k}_\perp 
= (k_0, k_1)$ provided k is replaced by $({\bf k}^2 {\bf k}^2_\perp)^{1/4}$.}, 
and $A(ka),B(ka), C(ka)$ and
$\nu(ka)$ are known functions of $ka$ (for their power series expansions,
see e.g. Ref.\cite{dasghosh}) satisfying the `unitarity' relation
$B^2(ka) = A^2 (ka) + C^2 (ka)$. The parameter $a$ is the same as in 
Eqs.(\ref{eq:bone}, \ref{eq:bthree}). 
In Eq.(\ref{eq:bthirteeen}), the variables $w,\bw$ are related to the 
coordinates $u,\bu$ by
\be
au = e^{-w} \qquad {\rm and} \qquad a\bu = e^{-\bw},
\label{eq:bfourteen}
\ee
and $H^{(i)}\left(\nu,w+{i\pi / 2} \right)$ for $i = 1,2$ are the
associated Mathieu functions of the third and the fourth kinds, respectively. 

We also need the normalized wave functions $\Psi_{\bf k}(u)$. After Wick
rotation to the Euclidean signature, they are given by
\be
\Psi_{\bf k} (u) =  N(ka)~e^{-i{\pi \over 2}(\nu+1)} 
~{1\over u^2}H^{(2)} \left(\nu,-w - {i \pi \over 2} \right),
\label{eq:fifteen}
\ee
where $N(ka)$ is a normalization factor, which is again a power series
in $(ka)$ and whose low-$ka$ behavior is known \cite{dasghosh}. 
It is now possible to extract the one-point correlator $\langle \cO({\bf k})
\rangle_\bu$ using Eq.(\ref{eq:beight}). For our purposes, it suffices
to analyze Eq.(\ref{eq:beight}) in two asymptotic regimes -- 
$a \bu \ll 1$ and $ a \bu \gg 1$.

For $a\bu \ll 1$, the associated Mathieu functions
asymptote to the Hankel functions:
\be
H^{(1, 2)} \left(\nu,\bw+{i\pi\over 2} \right)
\quad \longrightarrow \quad H^{(1, 2)}_\nu \left(i { k \over \bu} \right),
\ee
where $H^{(1, 2)}_\nu (z)$ are Hankel functions of the first and the
second kinds, respectively. As in the case of $AdS_5 \times S^5$, at
$u \rightarrow \infty$, subtraction of the flat space Green function is 
irrelevant and the one-point function reduces precisely to the $AdS_5$ result:
\be
\Big< \cO({\bf k}) \Big>_\bu \,\,\, = \,\,\, \left({k \over \bu}\right)^2
~K_2 \left({k \over \bu} \right) + \cdots
\qquad \quad \,\,\, {\rm for} \,\,\, \qquad \quad
\bu \ll {1 \over a}. 
\label{eq:sixteen}
\ee

To find the `tomographic' size of the hologram of the bulk source, we 
perform Fourier transform of Eq.(\ref{eq:sixteen}) over the momenta ${\bf k}$:
\bea
\Big< \cO(\Delta x ) \Big>_\bu &=& \int {d^2 {\bf k} \over (2 \pi)^2}
e^{ - i {\bf k} \cdot \Delta x}
\Big< \cO({\bf k}) \Big>_\bu
\nonumber\\
&=& {4 \over \pi} {\bu^{-4} \over
[ (\Delta x)^2 + \bu^{-2}]^3},
\eea
yielding a tomograph identical 
to the $AdS_5$ result, Eq.(\ref{eq:beleven}).
Thus, when the source is located deep inside the bulk where the space-time 
asymptotes to $AdS_5 \times S^5$, one finds 
`tomograph' of the one-point correlator is such that the size of the hologram 
at the boundary exhibits UV/IR duality:
\be
\Big( \vert \Delta x \vert \Big)_0 \quad \sim \quad
{1 \over \bu} \qquad \quad {\rm for} \qquad \quad
\bu \ll {1 \over a}. 
\label{sixteenhalf}
\ee

For $a\bu \gg 1$, the `tomograph' would be quite different as it covers
the region close to the boundary, where the supergravity mode in question
practically perceives flat space-time. In this region, the Euclidean
bulk Green function becomes, for $\bu < u$,
\be
\cG_{\rm E} (u,\bu; {\bf k}) \,\, = \,\, \cG_0 (u,\bu; {\bf k}) \, - \,
{1\over 2ka^2}{1\over (u\bu)^{5/2}}~{\hat{B} (ka) \over i A(ka)}~
e^{-ka^2 (u + \bu)} + \cdots,
\label{eq:seventeen}
\ee
where $\hat{B}(ka)$ refers to the real part of $B(ka)$.
Thus, the one-point correlator in this region is given by
\be
\Big< \cO({\bf k}) \Big>_\bu \,\, = \,\, 
\left( {\pi\over 8 ka^2 \bu^5} \right)^{1/2} {i\hat{B}(ka)\over N(ka)
A(ka)}~e^{-ka^2 \bu} + \cdots \qquad {\rm for} \qquad
\bu \gg {1 \over a}.
\label{eq:eighteen}
\ee
As anticipated, dependence on the cutoff $u$ has cancelled out completely 
so that the limit in Eq.(\ref{eq:beight}) indeed exists. 
From the form of power series expansion of the known coefficients 
$A(ka), B(ka), C(ka)$ and $N(ka)$ \cite{dasghosh}, one finds that
\be
{iB(ka)\over N(ka) A(ka)} \,\, = \,\, \left(-{2 \over 3a^2} \right) 
\Big[
1 + \alpha_1 (ka)^4 + \alpha_2 (ka)^4 \log ka + \cdots \Big].
\ee
Here, $\alpha_1, \alpha_2, \cdots$ are calculable numerical coefficients.
To tomograph the hologram of the supergravity 
source on the boundary, we perform 
Fourier transform of Eq.(\ref{eq:eighteen}) over the momentum ${\bf k}$:
\be
\Big< \cO( \Delta x) \Big>_\bu 
\,\,\, = \,\,\, -{\sqrt{\pi} \over 3a^2} \int {d^2 {\bf k} \over (2\pi)^2}
~e^{-i{\bf k}\cdot \Delta x}~
{1\over (2 k a^2 \bu^5)^{1/2}}~e^{- k a^2 \bu}~\Big[ 1 + \alpha_1 (ka)^4
+ \cdots \Big],
\label{eq:bnineteen}
\ee
The result is
\be
\Big< \cO(\Delta x) \Big>_\bu \,\,\, = \,\,\,
- {(a \bu)^{5/2} \over 12 \sqrt{2}a^2}~ \left[1 + 
\alpha_1 a^{-4} \partial^4_\bu + \cdots \right] 
P_{1/2}(Z(\bu)) ~\left( {Z(\bu) \over a \bu} \right)^{3/2}
\qquad {\rm for} \qquad \bu \gg {1 \over a}.
\label{eq:twenty}
\ee
Here, $P_\nu (Z)$ denotes the Legendre function and
$Z(\bu) = ( 1 + (\Delta x / a^2 \bu)^2)^{-1/2}$.
It is easily seen that the profile drops to zero fast for
$|\Delta x| \gg a^2 \bu$.  The characteristic size of the hologram can be
determined from equal-altitude contours of Eq.(\ref{eq:twenty}) being
a multiple of $\bu^2$ --- intersection locus of $P_{1/2}(Z)$ and 
$Z^{-3/2}$. The result exhibits UV/IR {\sl proportionality}:
\be
\Big(\vert\Delta x \vert \Big)_0 \quad \sim  \quad a^2 \bu
\qquad \quad {\rm for} \qquad \bu \gg {1 \over a}.
\label{eq:twentyone}
\ee
Thus, in this regime, the hologram size actually {\em increases} as the 
source moves closer to the boundary. If we impose an infrared cutoff $u_0$ in 
the supergravity background, then a source placed at the cutoff would 
correspond to a state in the gauge theory in which the dual operator is spread
over a size $a^2 u_0$ \footnote{This UV/IR proportionality between the size
and the location was also observed in \cite{danielsson} in the context
of holography in full D3-brane geometry.}.

The result is certainly consistent with the fact that gauge-invariant 
open Wilson line operators in the noncommutative gauge theory have the 
property that their size increases with increasing momentum. The following
heuristic picture may provide yet another way of understanding the UV-IR
proportionality. Consider Fourier-transforming the open Wilson line 
operator, Eq.(\ref{openwilson}), over ${\bf k}$-momenta with an
{\sl ultraviolet} cutoff $\Lambda$ and construct a position space 
`tomograph'. This means that, due to Eq.(\ref{uvir}), we are fixing the base 
point $x^\mu$ of the Wilson line and integrating over all possible separations 
$\Delta y^\mu$ --- the Fourier integral with some ultraviolet cutoff
$\Lambda$ would involve a sum over such Wilson lines with sizes
upto $\theta \Lambda$. Since the infrared cutoff of the bulk is
related to the ultraviolet cutoff of the boundary theory
by $u_0 \sim \Lambda$, one is summing over open Wilson lines of size up to 
$\theta u_0$. If we take into account of the strong-coupling effect that the 
true noncommutativity scale following from the dual supergravity differs from 
naive gauge theory estimate $\theta$ by a factor of $R$, the maximal size
of the open Wilson lines is the same as predicted by the dual supergravity,
viz $a^2 u_0$.

\begin{figure}[ht]
   \vspace{0.5cm}
\centerline{
   {\epsfxsize=6cm
   \epsfysize=4.5cm
   \epsffile{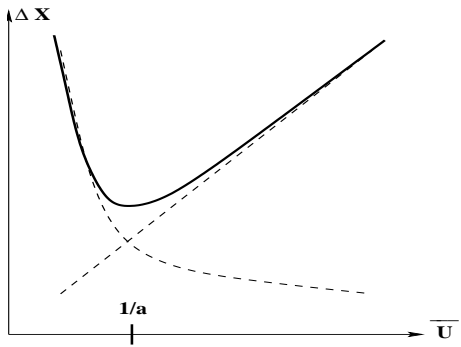}}
}
\caption{\sl Schematic view of the UV-IR relation in noncommutative
gauge theory.}
\label{uvirgraph}
\end{figure}

Combining the two regimes -- Eq.(\ref{sixteenhalf}) at low-energy regime
and Eq.( \ref{eq:twentyone}) at high-energy regime, we conclude that
the supergravity prediction for the UV/IR relation agrees well with 
that in noncommutative gauge theory, and takes a form depicted in 
Fig.(\ref{uvirgraph}). The agreement may be taken as a convincing 
evidence for holographic dual relation between the supergravity in
the background Eqs.(\ref{eq:bone}, \ref{eq:bthree}) and the noncommutative
gauge theory.  

The open Wilson line operator along a given contour and with definite momentum
would be, in general, dual to a combination of various supergravity fields
with the same momentum but with various quantum numbers. To understand the 
detailed operator-field relationship, one needs to learn how to decompose 
the Wilson line operator irreducibly according to the quantum numbers of 
supergravity modes. We will address this issue in a separate publication. 
It should be noted, however, that the IR/UV properties demonstrated here are 
quite independent of the details of such a decomposition. 

Finally, it would be also interesting to extend the methods we have developed
in this paper to the noncritical open string theories \cite{ncos}
which arises as a decoupling limit in open string theory with `electric' 
noncommutativity and to the dynamics of the noncommutative monopoles,
fluxons \cite{gross} and vortices \cite{vortices}. 
We hope to return this question in future publications.

\section*{Acknowledgement}
We thank P. Kraus and S. Mathur for discussions. SRD acknowledges warm 
hospitality of School of Physics at Seoul National University, where this
work was initiated. SJR acknowledges warm hospitality of Theory Division 
at CERN, where this work was completed.

\end{document}